# Spontaneous non-stoichiometry and ordering of metal vacancies in degenerate insulators

Oleksandr I. Malyi[1,2], Michael T. Yeung[3], Kenneth R. Poeppelmeier[3], Clas Persson[2], Alex Zunger[1*]

[1]Renewable and Sustainable Energy Institute, University of Colorado, Boulder, Colorado 80309, USA

[2]Centre for Materials Science and Nanotechnology, Department of Physics, University of Oslo, P.O. Box 1048 Blindern, NO-0316 Oslo, Norway

[3]Department of Chemistry, Northwestern University, Evanston, Illinois 60208, USA

[*]Corresponding author
e-mail: Alex.Zunger@Colorado.edu

## SUMMARY

We point to a class of materials representing an exception to the Daltonian view that compounds maintain integer stoichiometry at low temperatures, because forming stoichiometry-violating defects cost energy. We show that carriers in the conduction band (CB) of degenerate insulators in Ca-Al-O, Ag-Al-O, or Ba-Nb-O systems can cause a self-regulating instability, whereby cation vacancies form exothermically because a fraction of the free carriers in the CB decay into the hole states formed by such vacancies, and this negative electron-hole recombination energy offsets the positive energy associated with vacancy bond breaking. This Fermi level-induced spontaneous non-stoichiometry can lead to the formation of crystallographically ordered vacancy compounds (OVCs), explaining the previously peculiar occurrence of unusual atomic sequences such as $Ba_lNb_mO_n$ with l:m:n ratios of 1:2:6, 3:5:15, or 9:10:30. This work has general ramifications as the degenerate insulators have found growing interests in many fields ranging from transparent conductors to electrides that are electron donating promotors for catalyst.

# INTRODUCTION: Spontaneous non-stoichiometry as a Fermi level instability

The fact that compounds manifest *integer* ratios between component elements ("the law of definite proportions"[1]) has been the cornerstone of our understanding of formal oxidation states (taking up integer values) and defect physics (showing that violation of integer ratios by formation of defects costs energy and is thus unlikely at low temperatures). This thinking of the *daltonides* school (the paradigm of stoichiometry) stood in stark contrast with the non-stoichiometric *berthollides*[2] school, who argued that compounds could possess a range of compositions entirely dependent on the starting synthetic conditions. We point out an interesting class of exceptions to the Daltonide universal understandings, whereby a degenerate insulator with Fermi energy ($E_F$) inside the conduction band and a large "internal band gap" ($E_g^{int}$) between the valence band maximum (VBM) and conduction band minimum (CBM) (depicted in Fig. 1a) could form significant concentration of low-energy vacancies, violating the rule of integer stoichiometry. Compounds with degenerate conduction bands and an internal VBM-to-CBM band gap that manifest peculiar cation non-stoichiometry are common (but generally unrecognized as a generic phenomenon), and include chemically diverse classes such as main-group oxides $Ca_6Al_7O_{16}$[3,4], noble metal oxides $Ag_3Al_{22}O_{34}$[5], as well as transition metal oxides such as $Sr[V,Nb]O_3$[6,7], $CaVO_3$[6], and $Ba_3[Nb,Ta]_5O_{15}$.[8-10] Interestingly, many of these compounds were also proposed[5,6] to be "transparent conductors", thereby combining the generally contradictory attributes of metallic conductivity with insulator-like transparency. We explain this phenomenon of spontaneous low-temperature non-stoichiometry (violation of the ideal *daltonides*, but consistent with *berthollides*) by the energy gain associated with the transfer of high-energy conduction band electrons to the low-energy defect level ("DL" in Fig. 1a, such as metal vacancy electron acceptor state). We show, on general grounds, via the first-principles defect theory[11], that degenerate insulators with sufficiently large internal gap can have in the dilute vacancy limit the unusual *negative vacancy formation enthalpy*, thus violating the standard textbook notion that formation of defects costs energy and is thus unlikely at low temperatures. At the concentrated limit, such vacancies can condense into ordered crystallographic arrays, thus explaining the hitherto peculiar occurrence[12] of macroscopically observed sequences of ground state OVCs[13-16], such as $Ca_lAl_mO_n$ with l:m:n ratios of 11:14:32 and 23:28:64, or $Ba_lNb_mO_n$ with l:m:n ratios of 1:2:6, 3:5:15, 5:4:15, 7:6:21, 7:8:24, 9:10:30, and 26:27:81. These otherwise surprising integer ratios are obtained here as stable, T=0 K ground state structures via first-principles total energy search for OVC compositions and geometries that are on the ground state line ("convex hull").

What makes this mechanism for spontaneous generation of vacancies interesting, in addition to its extraordinary violation of Dalton's law of precise stoichiometry, is the fact that it has a major impact on transparent conducting oxides (TCOs), and more broadly to electrides[3, 17], tungsten bronzes[7, 9, 18, 19], and generic *low temperature spontaneous non-stoichiometry*. The former class of materials represents an exceptional resolution of otherwise mutually exclusive properties: while metallic conductors are generally opaque, whereas transparent compounds are generally electrical insulators, a small group of compounds exhibits reasonably good transparency and sufficiently high conductivity.[20] Unfortunately, controlling conductivity and transparency in TCOs is a major challenge largely because we do not know of a "knob" that decouples the transparency and conductivity in a compound. Interestingly, since different l:m:n values in OVCs deplete the conduction band of electrons by different amounts, thus reducing unwanted plasma absorption/reflectivity, while diminishing the wanted conductivity, control of l:m:n during growth is the requisite knob for control of TCOs. Interestingly, different l:m:n OVCs are predicted to be thermodynamically stable in different ranges of chemical potentials of the constituent elements, so in principle selection of a given OVC can be controlled via synthesis. This provides a knob for control of transparency and conductivity. We conclude that cation vacancies created as a result of populating the conduction band by electrons are agents that mitigate the contradictory properties upon which transparent conductors are fundamentally based.

We validate our theory with two key experimental findings: silver beta-alumina $Ag_3Al_{22}O_{34.5}$ leeches out silver atoms upon any attempt to introduce electron carriers, and the tetragonal tungsten bronze $Ba_3Nb_5O_{15}$ where the cation vacancy formation favors the formation of secondary phases. Our paper is the first to attempt to understand and unravel *the intrinsic complex connection between stoichiometry and properties,* creating a roadmap for the future. This work has broad implications as degenerate insulators with such unique band structures have been generating increasing interest in many fields beyond that of transparent conductors including the colored metallic photocatalysts as seen in the substoichiometric $Sr_{1-x}NbO_3$[7], the electron donating promotors for catalysts[21] to low work function compliance layers.[22] Furthermore, our work explains why many of these degenerate insulators require exotic synthesis conditions to be prepared stoichiometrically.

## RESULTS AND DISCUSSION

### Dilute metal vacancies can form readily in degenerate insulators

The formation of vacancies in non-degenerate insulators (Fig. 1b) involves breaking of stable chemical bonds without restoring any of the spent energy, so a significant concentration of such defects can exist only at elevated temperatures. The situation can be different in degenerate insulators (with $E_F$ located at an energy $\Delta E_{CB}$ above the CBM, as in $BaNbO_3$ and $Ca_6Al_7O_{16}$ discussed below) having an "internal band gap" $E_g^{int}$ between the CBM and the VBM (Fig. 1a). Here, the formation of dilute metal vacancy can create electron acceptor states near the valence band (at energy $E_{DL}$), resulting in the opportunity for decay of $q$ conduction band electrons into these electron acceptor states, thereby regaining the energy $q(E_g^{int}+\Delta E_{CB}-E_{DL})$ which reduces accordingly the vacancy formation energy (see Fig. 1a).

To validate this concept, Figs. 2a, c, e show the results of density-functional calculated formation energies of the Ba vacancy in $BaNbO_3$, the Ca-vacancy in $Ca_6Al_7O_{16}$, and the Ag vacancy in $Ag_3Al_{22}O_{34}$ systems as a function of the metal chemical potential (see Methods). The allowed stable chemical potential regions (constructed by considering possible competing phases, see Methods) of the respective bulk compounds are shown in Figs. 2b, d, f. We see that for degenerate insulators under cation deficient chemical potential conditions, vacancy formation energies can be extremely low (in fact, negative). Whereas $BaNbO_3$ and $Ca_6Al_7O_{16}$ have stable chemical potential (green) zones at the respective stoichiometries indicated, $Ag_3Al_{22}O_{34}$ does not. In fact, in the latter case the Ag vacancy formation (Fig. 2e) is so strongly negative under all chemical potential conditions, that the conduction band is empty and the parent degenerate $Ag_3Al_{22}O_{34}$ phase is not stable (i.e., no green zone in Fig. 2f).

### Dilute vacancies can condense into stable ordered vacancy compounds forming homological sequences

The negative formation energies of *dilute* vacancies open the possibility of vacancy condensation and long-range ordering. To examine this possibility, we have calculated the T=0 K stable phases ("ground state diagram" or "convex hull") of such ternary structures. This entails searching for configuration versus composition that lies on the energy convex hull[23] which defines the phases having energy lower than a linear combination of any competing phases at the corresponding compositions. We create candidate configurations by considering a base compound ($BaNbO_3$, $Ba_3Nb_5O_{15}$, $Ca_6Al_7O_{16}$, or $Ag_3Al_{22}O_{34}$), then create a replica of $N$ such units of base compounds and add successively $p$ metal vacancies, i.e.,

OVC=$N\times$(base)+$p$V$_m$, searching via total energy minimization for stable and metastable configurations. The results are summarized in Fig. 3. *The key points to notice is that there is a sequence of OVC phases, each having a specific concentration of electrons per formula unit (e/f.u) in the conduction band (including possibly zero)*. This is illustrated in Figs. 4a and 5a showing the ground state diagrams for Ba-Nb-O and Ca-Al-O systems and chemical potential stability diagrams (Figs. 4b and 5b) demonstrating the phases that are stabilized as the chemical potentials of the atoms being removed are continuously changed between their allowed values.

***Stable phases and OVCs for the Ba-Nb-O system:*** we find 25 binary and *ternary ground state compounds* (described in Supplementary Data 1) of which 7 are OVCs (indicated in the following in boldface): $NbO$, $NbO_2$, $Nb_2O_5$, $Nb_8O$, $Nb_{12}O_{29}$, $BaO$, $BaO_2$, $Ba_2Nb_5O_9$, $BaNb_7O_9$, $Ba_6Nb_2O_{11}$, $BaNb_5O_8$, $Ba_4Nb_2O_9$, $BaNb_4O_6$, $Ba_2Nb_{15}O_{32}$, $BaNb_8O_{14}$, $Ba_4Nb_{14}O_{23}$, $Ba_3Nb_{16}O_{23}$, $BaNbO_3$, **$Ba_7Nb_6O_{21}$**, **$Ba_5Nb_4O_{15}$**, **$Ba_3Nb_5O_{15}$**, **$Ba_7Nb_8O_{24}$**, **$Ba_9Nb_{10}O_{30}$**, **$Ba_{26}Nb_{27}O_{81}$**, and **$BaNb_2O_6$**. The compounds 1:2:6, 3:5:15, and 5:4:15 OVCs were observed experimentally before.[10, 24, 25] However, not all predicted ground state compounds are currently known. For example, $Nb_8O$, $Ba_6Nb_2O_{11}$, $Ba_4Nb_2O_9$, $Ba_7Nb_8O_{24}$, $Ba_7Nb_6O_{21}$, and $Ba_{26}Nb_{27}O_{81}$ structures have never been reported, and constitute opportunities for novel synthesis.

***Stable phases and OVCs for the Ca-Al-O system:*** We identify a total of 10 binary and ternary *ground state compounds* of which two are OVCs (indicated in what follows in boldface) (described in Supplementary Data 2): $Ca_8Al_3$, $CaAl_2$, $CaAl_4$, $CaO_2$, $CaO$, $Al_2O_3$, $CaAl_4O_7$, $Ca_6Al_7O_{16}$, **$Ca_{11}Al_{14}O_{32}$**, and **$Ca_{23}Al_{28}O_{64}$**.

***Stable phases and no OVCs for the Ag-Al-O system:*** We find 4 binary and ternary *ground state compounds* (described in Supplementary Data 3): $Al_2O_3$, $Ag_2O$, $AgAl$, and $AgAlO_2$. None of them are OVCs. In fact, the $Ag_3Al_{22}O_{34}$ and its $Ag_2Al_{22}O_{34}$ OVC lie energetically above the convex hull by 0.034 and 0.01 eV/atom, respectively. Specifically, $Ag_3Al_{22}O_{34}$ decomposes to $AlAgO_2$, $Ag$, and $Al_2O_3$ phases, while $Ag_2Al_{22}O_{34}$ decomposes to $AgAlO_2$ and $Al_2O_3$. In other words, the formation of Ag vacancy is highly effective and drains all electrons from conduction band, thus $Ag_2Al_{22}O_{34}$ is an insulator. This vacancy formation increases the system stability (the energy above the convex hull is smaller for $Ag_2Al_{22}O_{34}$ than that for $Ag_3Al_{22}O_{34}$), but both $Ag_3Al_{22}O_{34}$ and $Ag_2Al_{22}O_{34}$ are unstable with respect to competing phases. The results suggest that despite the fact that $Ag_3Al_{22}O_{34}$ might be attractive as an intrinsic transparent conductor[5] if Ag vacancy formation can be partially inhibited, the system is not likely to be realized experimentally under normal conditions as is also confirmed by our experimental results described in Supplementary Note I.

# Different ordered vacancy compounds have different number of free carriers, absorption, and conductivity

A. *Trends in the number of electrons in the conduction band of ITCs and the metal-insulator transition*

**For Ba-Nb-O (Fig. 3)**, among the ground state structures, the phases 1:1:3, 7:6:21, 26:27:81, 9:10:30, 7:8:24, and 3:5:15 have electrons in the CB and wide internal band gaps. From the sequences of OVCs, we find that the both Ba and Nb vacancies act as acceptors, removing 2$e$ and 5$e$ per vacancy from the conduction band (Fig. 4c). The electronic properties of Ba-Nb-O systems were studied experimentally by various groups.[9, 10, 18, 19, 26-28] Specifically, it has been shown that the 1:1:3 compound is a metal[19], while the 1:2:6 and 5:4:15 OVCs are insulators.[27, 28] At the same time, experimental results on electronic properties of the 3:5:15 compound from different studies differ from each other. In particular, despite a few works reporting low resistivity for the 3:5:15 compound, the temperature dependence of the resistivity (metal versus semiconductor nature) is different for different studies.[9, 10, 18, 26] These results are not surprising considering the tiny range of chemical potentials under which the 3:5:15 OVC exists (Fig. 4b), low defect formation energy in degenerate insulators, and complexity of material synthesis (Supplementary Note II). Furthermore, the sensitivity of sample preparation conditions reignites the need for high-quality single crystals to quantify potential vacancy concentration and to understand the structural/electronic impact of vacancy formation.

**For Ca-Al-O (Fig. 3)**, from the analysis of electronic properties of the ground state compounds in the Ca-Al-O system, we identify the phases 6:7:16 and 23:28:64 as degenerate insulators having 1$e/f.u$ and 2$e/f.u$ in the conduction band, respectively. At the same time, the 11:14:32 OVC has no conduction band electrons, i.e., it is an insulator with a wide band gap. The results for electronic properties of OVCs suggest that Ca vacancy in 6:7:16-based degenerate insulators acts as acceptor removing 2$e$ from the conduction band. However, in contrast to degenerate insulators in the Ba-Nb-O system, the vacancy formation results in the formation of in-gap occupied defect states which reduces both the internal band gap energy and energy range of occupied states (Fig. 5c). The high electronic conductivity of 6:7:16 is known from experimental studies[3, 4] while insulating and degenerate insulating properties of 11:14:32 and 23:28:64 OVCs are yet to be confirmed. It should be also noted that 6:7:24 and its 23:28:64 OVC have clearly defined 0-dimensional charge carrier density localization which implies that both compounds are not only degenerate insulators but also stable inorganic electrides (Supplementary Note III).

**B. *Spontaneously formed vacancies enhance transparency while reducing conductivity***

The optical properties of a degenerate insulator are determined by superposition of interband and intraband transitions (Fig. 6a). Owing to the occupied conduction band, the band-to-band transitions can be divided on (I) interband transitions from occupied valence to unoccupied conduction bands and (II) interband transition from occupied conduction bands to higher energy unoccupied bands. In the simplified model used here, the intraband transitions can be predicted within the Drude model[29] describing free-electron absorption.

The contribution of the aforementioned different types of transitions to the absorption spectra is illustrated in Fig. 6b for BaNbO$_3$. Specifically, the interband transitions (I) contribute to the absorption spectra at energies slightly above $E_g^{int}$ +$\Delta E_{CB}$ due to the curvature of the valence band. The interband transitions (II) contribute noticeably at energies below those for interband transitions (I) and determine the absorption spectra in the range from 2-4 eV. Finally, the intraband contribution determines the low energy region of absorption spectra. This analysis defines the design principles for good ITCs that require (a) metals having large internal gaps between the valence and conduction bands to minimize the interband absorption in visible range and (b) a high enough carrier density ($n_e$) in the conduction band to provide conductivity, while having (c) a low enough carrier density, limiting the interband transition in conduction band and plasma frequency ($\omega_p \sim \sqrt{n_e/m}$, where $m$ is an effective carrier mass) so that free-electron absorption does not impede the needed optical transparency. Furthermore, (d) the free carriers in the conduction band above the internal band gap should not destabilize the compound by spontaneously creating Fermi-level induced defects that defeat conductivity.

*For Ba-Nb-O*, the 1:1:3 degenerate insulator has the highest plasma frequency ($\omega_p$= 4.43 eV) among the considered materials which results in strong Drude contribution in the visible light range. Hence, despite high carrier concentration ($n_e$= 1.38×10$^{22}$ cm$^{-3}$) and stability, the 1:1:3 compound is not attractive as a ITC due to low transparency. Indeed, the 1:1:3 samples have clearly distinct color in experimental studies.[19] At the low vacancy concentration, the effect of the defects on carrier density and optical properties is negligible. However, increase of Ba vacancy concentration and formation of OVCs noticeably change the optoelectronic properties via reducing plasma frequency contribution and modifying the material band structure. Although 26:27:81, 9:10:30, and 7:8:24 OVCs have smaller carrier concentrations compared to 1:1:3 compound, the free-electron absorption is still a limiting factor as shown on the example of 7:8:24 OVC (Fig. 6c). Among the stable degenerate Ba-Nb-O insulators, 3:5:15 and 7:6:21 compounds have the smallest average absorption coefficients in the visible range (highest transparency).

**For Ca-Al-O,** the optical properties in the visible range in large part are determined by the interband transition from occupied conduction bands to unoccupied bands. Specifically, for the 6:7:16 compound, the carrier concentration of $2.26\times10^{21}$ cm$^{-3}$ results in the averaged plasma frequency of 1.68 eV, which does not induce noticeable free-electron absorption in the visible range. The interband transitions from the valence to the conduction bands are well separated in energy and contributes only above $E_g^{int} + \Delta E_{CB}$, which is over 4.4 eV (Fig. 4c). Because of this, the absorption in the energy range from 1.5 eV to below $E_g^{int} + \Delta E_{CB}$ is mainly determined by interband transitions in the conduction band, acting as a killer of materials transparency. The Ca vacancy formation removing 2$e$ from the conduction band per defect not only reduces the plasma frequency but also changes the interband transition which is illustrated in the absorption spectra of 6:7:16 and its OVCs (see Fig. 6d). In particular, 23:28:64 OVC has the lowest absorption in the visible light range among the considered degenerate insulators.

### C. *Stability of OVCs during growth is the knob that controls vacancy concentration hence transparency and conductivity*

We have seen above that different OVCs have a different number of free carriers, absorption, and conductivity. What makes this a useful feature is that *each such OVC can be realized in a unique and specific range of atomic chemical potentials that can, in principle, be controlled during growth.* This is illustrated in Fig. 4b for the Ba-Nb-O system and in Fig. 5b for the Ca-Al-O system, whereas the Ag-Al-O system offers no control as the Ag vacancy forms readily for any chemical potential. Significantly, the results in Figs. 4b, 5b suggest that it is possible to stabilize different Ba$_l$Nb$_m$O$_n$ and Ca$_l$Al$_m$O$_n$ OVCs by tuning Ba/Nb and Ca chemical potentials, respectively. As an illustration, considering the chemical potential stability zone for BaNbO$_3$ (Fig. 4b), we find that the reduction of the Ba chemical potential results in stabilization of the 26:27:81, 9:10:30, 7:8:24, 3:5:15, and 1:2:6 OVCs, while 7:6:21 and 5:4:15 OVCs are stable under Nb-poor condition. Formation of non-stoichiometric structures for Ba-Nb-O materials family has also been reported experimentally (e.g., Ba$_{0.95}$NbO$_3$)[30]; our results extend the knowledge on non-stoichiometric structures, showing how non-stoichiometry often seen in the system is an electronic effect – a high Fermi energy induces the formation of electron-killer cation vacancies. This effect is argued to be rather widespread, not an exotic curiosity, as there are many compounds with a large internal gap and Fermi level in the conduction band. For instance, fairly recently, Sr$_{1-x}$NbO$_3$ (a BaNbO$_3$- like metal whose Fermi level sits above a 1.8 eV gap) was found to be rare, colored, metallic photocatalyst[7] whose coloring and electronic conductivity can be controlled by synthesis conditions. Our results also provide insights into the stability of the transparent conductive states while pointing out that controllable formation of non-stoichiometric degenerate insulators can be used to design next-generation TCOs. Specifically, among the stable OVCs, $Ba_3Nb_5O_{15}$, $Ba_7Nb_6O_{21}$, and $Ca_{23}Al_{28}O_{64}$ OVCs are potential TCOs. However, $Ba_3Nb_5O_{15}$ exists only under an extremely tiny range of chemical potentials (see Fig. 4b). Because of this and low defect formation energy in degenerate insulators, the further research required to quantify the sensitivity of sample preparation conditions on the structural/electronic properties of the material. $Ba_7Nb_6O_{21}$ and $Ca_{23}Al_{28}O_{64}$ OVCs are first predicted to be stable ground state compounds and constitute opportunities for novel synthesis. Here, $Ca_{23}Al_{28}O_{64}$ OVC has the highest transparency among all considered systems, but due to small stability zone, a precise control of chemical potentials can be needed to realize the compound. $Ba_7Nb_6O_{21}$ exists in the widest range of chemical potentials among potential TCOs which implies simpler control of synthesis conditions to realize the material.

## CONCLUSIONS

The existence of non-stoichiometry in oxides is often thought to be a growth effect rather than a specific electronic instability. However, we demonstrate via first-principles calculations that degenerate insulators with sufficiently large internal gap and Fermi level in conduction band can have a characteristically negative cation vacancy formation enthalpy; in other words, spontaneous non-stoichiometry occurs even at low temperatures, intrinsic to the compound (not due to extrinsic effects). At the concentrated limit, such vacancies condense into OVCs. We show that this is a generic behavior as cation vacancy formation results in the decay of conduction band electrons into electron acceptor states. As a result, the negative electron-hole recombination energy offsets the positive energy associated with vacancy bond breaking. Our results thus explain and clarify how non-stoichiometry often seen in oxides is an electronic effect – a high Fermi energy induces the formation of electron-killer acceptors. This effect is argued to be rather widespread, not an exotic curiosity, as there are many compounds with such electronic structures. We demonstrate by the chemical potential phase diagrams the growth regimes where a particular l:m:n OVC with a fixed level of transparency and conductivity is stable. For example, whereas in $Ag_3Al_{22}O_{34}$ the Ag vacancies form readily (so the conduction band is completely depleted of carriers), in the Ca-Al-O and Ba-Nb-O systems there are well-defined chemical potential domains, where OVCs with an optimal number of carriers persist. Since each l:m:n OVC depletes the conduction band of electrons by a different amount, the controllable formation of OVCs can be used to modify interband absorption, enhance materials stability, reduce plasma absorption, and design next-generation intrinsic transparent conductors. Specifically, we identify $Ba_3Nb_5O_{15}$, $Ba_7Nb_6O_{21}$, and $Ca_{23}Al_{28}O_{64}$ OVCs as an attractive candidature to TCO application. This work has broad implications as degenerate insulators have been attracting much interest in many fields beyond that of transparent conductors.

**MATERIALS AND METHODS**

*(a) Density functional calculations:* All spin-polarized calculations are carried out using Perdew-Burke-Ernzerhof (PBE) functional[31] with DFT+U correction for Nb (U = 1.5 eV) and Ag (U = 5 eV) d-like orbitals as implemented by Dudarev *et al.*[32] and available in Vienna Ab Initio Simulation Package (VASP).[33] It should be noted that the utilization of DFT+U with larger U value (e.g., U = 3 eV) for Nb d-like orbitals results in incorrect insulating nature of $BaNbO_3$ system. The cutoff energies for plane-wave basis are set to 500 and 600 eV for final static and volume relaxation calculations, respectively. Γ-centered Monkhorst-Pack k-grids[34] are used in the Brillouin-zone integrations with the grid densities of approximately 2500 and 10000 per atom for volume relaxation and final static calculations. For each system, random atomic displacements within 0.1 Å are applied to avoid trap at a local minimum. The full optimization of lattice vectors and atomic positions is allowed. The obtained results are analyzed using Vesta[35] and pymatgen.[36]

*(b) Calculations of chemical potential domains:* To calculate stability zones for $BaNbO_3$, $Ca_6Al_7O_{16}$, and $Ag_3Al_{22}O_{34}$ compounds presented in Fig. 2, we use only experimentally known stoichiometric crystal structures available in Inorganic Crystal Structure Database (ICSD)[12] and SpringerMaterial.[37] To predict convex hulls and chemical potential diagrams for Ca-Al-O and Ag-Al-O systems, in addition to the known experimental structures, we include generated OVCs and structures available in Materials Project[38] database. For Ba-Nb-O (Figs. 4a, b), we also consider stoichiometries inspired by II-V-O phase diagrams. $O_2$ molecule is used as a reference state for $O_2$ phase. The ground state compounds found in this work are given in Supplementary Data I-III. We implement fitted elemental-phase reference energies (FERE) corrections[39] to correct chemical potentials for elemental reference states.

*(c) Optical spectra:* The optical properties (Fig. 6) are computed within independent particle approximation.[40] To calculate plasma frequencies and interband transition spectra, 40×40×40, 20×20×20, 20×20×20, 8×24×8, 16×12×8, 20×20×20, 8×8×8, and 8×8×8 Γ-centered k-point grids are used for $BaNbO_3$, $Ba_7Nb_8O_{24}$, $Ba_7Nb_6O_{21}$, $Ba_3Nb_5O_{15}$, $BaNb_2O_6$, $Ca_6Al_7O_{16}$, $Ca_{23}Al_{28}O_{64}$, $Ca_{11}Al_{14}O_{32}$, respectively. The Drude contribution to optical properties is included by utilizing kram code[41] with plasma frequencies calculated from first-principles calculations and the damping coefficient of 0.2 eV, which is analogous to traditional TCs.[42]

*(d) Defect calculations:* 116, 135, and 236-atom supercells are used to calculate defect energetics in $Ca_6Al_7O_{16}$, $BaNbO_3$, and $Ag_3Al_{22}O_{34}$ systems, respectively. The defect formation energies (Figs. 2a, c, e) and finite size corrections are computed within the framework described by Lany and Zunger[43, 44] and implemented in the pylada-defects code.[45] The ranges of chemical potentials are determined using only experimentally known stoichiometric crystal structures as described above.

**Acknowledgment:** The work of A.Z. and O.M. at CU and of K.R.P. and M.T.Y. at NU was supported by NSF-DMR EPM program under grant DMR-1806939 and DMR-1806912, respectively. M.T.Y. would like to thank the Department of Energy (DOE) Office of Energy Efficiency and Renewable Energy (EERE) Postdoctoral Research Award under the EERE Solar Energy Technologies Office administered by the Oak Ridge Institute for Science and Education (ORISE) for the DOE. ORISE is managed by Oak Ridge Associated Universities (ORAU) under DOE contract number DE-SC00014664. All opinions expressed in this paper are the author's and do not necessarily reflect the policies and views of DOE, ORAU, or ORISE. OM and CP acknowledge the support from Research Council of Norway (contract: 251131), and Norwegian Metacenter for Computational Science (NOTUR) for providing access to supercomputer resources.

**Author contributions:** O.I.M. carried out theoretical calculations; M.T.Y. fabricated the samples and carried out structure analysis by XRD and TEM. A.Z. directed the design of the research, analysis of the results and the writing of the paper. O.I.M. contributed the most to the writing of the paper with contributions from all co-authors. K.R.P. supervised the experimental work. C.P. and A.Z. supervised all theoretical studies.

**Competing interests:** The authors declare no competing interests.

**Data Statement:** All data needed to evaluate the conclusions in the paper are present in the paper and the Supplementary Materials. Additional data related to this paper may be requested from the authors.

# SUPPLEMENTAL INFORMATION

Supplementary Note I: Experimental observation of Ag formation in $Ag_3Al_{22}O_{34+x}$ under reduced atmosphere

Supplementary Note II: Synthesis of $Ba_3Nb_5O_{15}$

Supplementary Note III: The delocalized nature of the conducting electrons in degenerate insulators

Supplementary Data I: Ground state compounds in Ba-Nb-O system

Supplementary Data II: Ground state compounds in Ca-Al-O system

Supplementary Data III: Ground state compounds in Ag-Al-O system

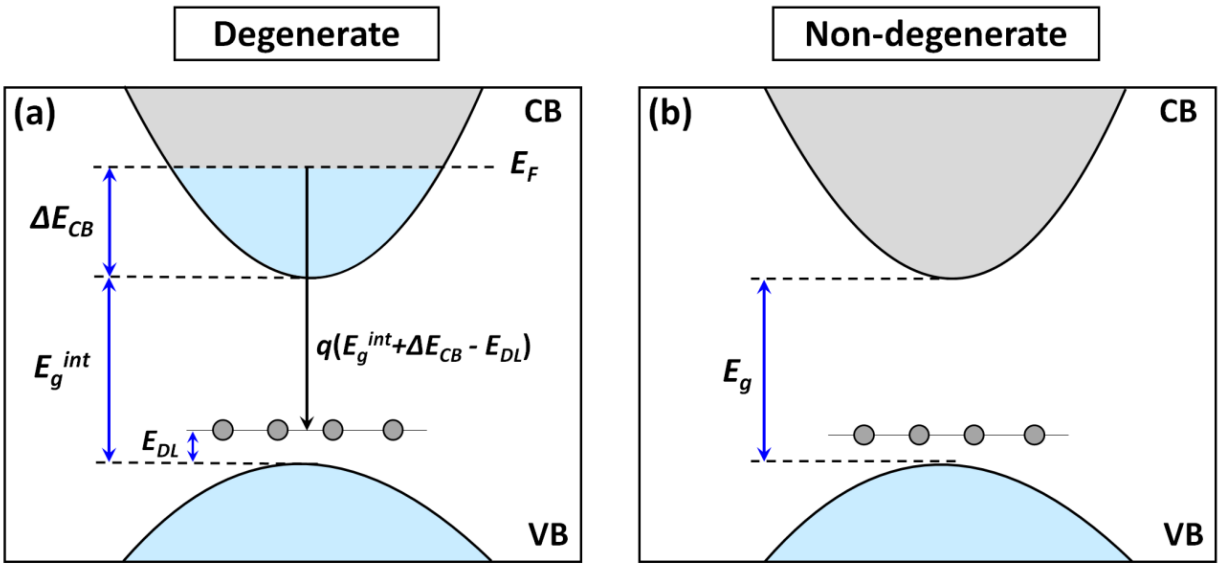

**Figure 1.** Schematic illustration of defect formation in (a) degenerate and (b) non-degenerate insulators showing lowering defect formation energy governed by removing an electron from the conduction band. blue: occupied states, grey: empty states. Here, $\Delta E_{CB}$ is the occupied part of the conduction band; $E_g^{int}$ is the internal band gap, and $E_{DL}$ is the electron-trap defect level produced by, e.g., cation vacancies.

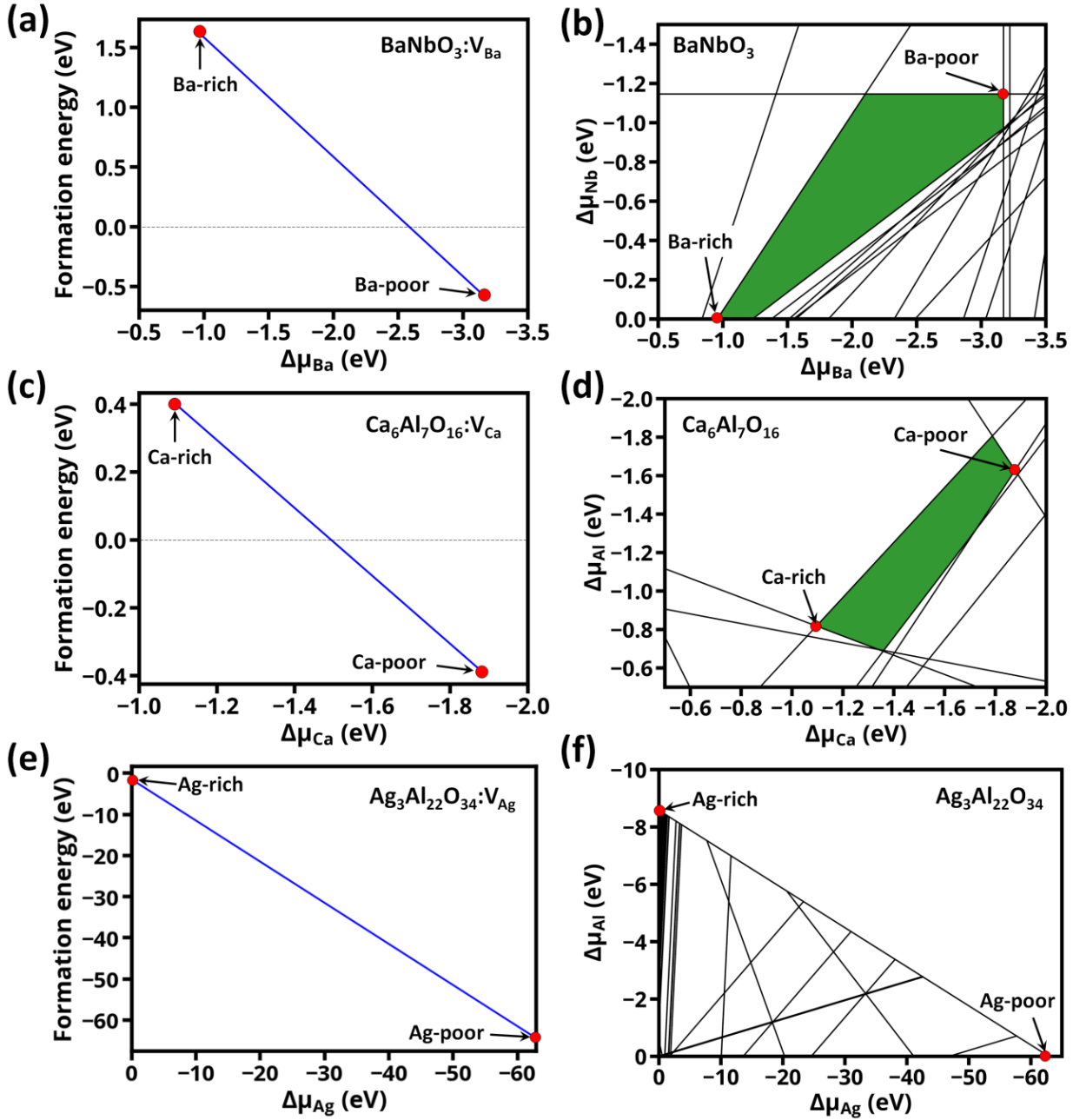

**Figure 2.** Metal vacancy formation energetics (left hand side panels) and regions of chemical potentials where the degenerate insulator is stable (right hand side panels) for $BaNbO_3$, $Ca_6Al_7O_{16}$, and $Ag_3Al_{22}O_{34}$. For $Ag_3Al_{22}O_{34}$, no green zone exists where the compound is stable.

| OVC generator | Resulting OVC | $N_e$ (e/f.u) | $\Delta E_{CB}$ (eV) | $E_g^{int}$ (eV) | Material type | Symmetry |
|---|---|---|---|---|---|---|
| Ba-Nb-O | | | | | | |
| 1:1:3 (Base) | $BaNbO_3$ | 1 | 1.11 | 2.26 | I | 1:1:3 |
| 27×(1:1:3)+$V_{Ba}$ | $Ba_{26}Nb_{27}O_{81}$ | 25/27 | 1.04 | 2.22 | I | 1:1:3-like |
| 10×(1:1:3)+$V_{Ba}$ | $Ba_9Nb_{10}O_{30}$ | 8/10 | 0.90 | 2.19 | I | 1:1:3-like |
| 8×(1:1:3)+$V_{Ba}$ | $Ba_7Nb_8O_{24}$ | 6/8 | 0.88 | 2.17 | I | 1:1:3-like |
| **7×(1:1:3)+$V_{Nb}$** | **$Ba_7Nb_6O_{21}$** | **2/7** | **0.40** | **2.20** | **II** | **Reconstructed** |
| **5×(1:1:3)+2$V_{Ba}$** | **$Ba_3Nb_5O_{15}$** | **1/5** | **0.46** | **1.87** | **II** | **Reconstructed** |
| 2×(1:1:3)+$V_{Ba}$ | $BaNb_2O_6$ | 0 | 0 | 2.72 | III | Reconstructed |
| 5×(1:1:3)+$V_{Nb}$ | $Ba_5Nb_4O_{15}$ | 0 | 0 | 2.65 | III | Reconstructed |
| Ca-Al-O | | | | | | |
| 6:7:16 (Base) | $Ca_6Al_7O_{16}$ | 1 | 0.96 | 3.49 | I | 6:7:16 |
| **4×(6:7:16)+$V_{Ca}$** | **$Ca_{23}Al_{28}O_{64}$** | **2/4** | **0.67** | **3.10** | **II** | **6:7:16-like** |
| 2×(6:7:16)+$V_{Ca}$ | $Ca_{11}Al_{14}O_{32}$ | 0 | 0 | 3.40 | III | 6:7:16-like |

**Figure 3**. Summary for results of stable ordered vacancy compounds (OVCs) in Ba-Nb-O and Ca-Al-O systems. OVCs are generated by creating supercells of a base cell as OVC=$N$×(base)+$p$$V_m$, where $N$ and $p$ are integer numbers. Number of electrons ($N_e$) in the conduction band is given per formula unit (f.u.) of the base compound. In general, the number of electrons in the conduction band for the degenerate insulators in Ca-Al-O and Ba-Nb-O systems can be predicted from the sum of composition-weighted formal oxidation states assuming $Ba^{2+}$, $Ca^{2+}$, $Al^{3+}$, $Nb^{5+}$, and $O^{2-}$ for each compound. $\Delta E_{CB}$ is the occupied energy range of conduction band (Fig. 1a) and $E_g^{int}$ is the internal gap between CBM and VBM (Fig. 1a) estimated from the density of states. Type of materials I, II, and III stands for not transparent metals, potentially transparent conducting oxides, and insulator, respectively. Potentially transparent conducting oxides are denoted in red. If the OVC in the lowest energy state is strongly reconstructed, it is labeled as reconstructed.

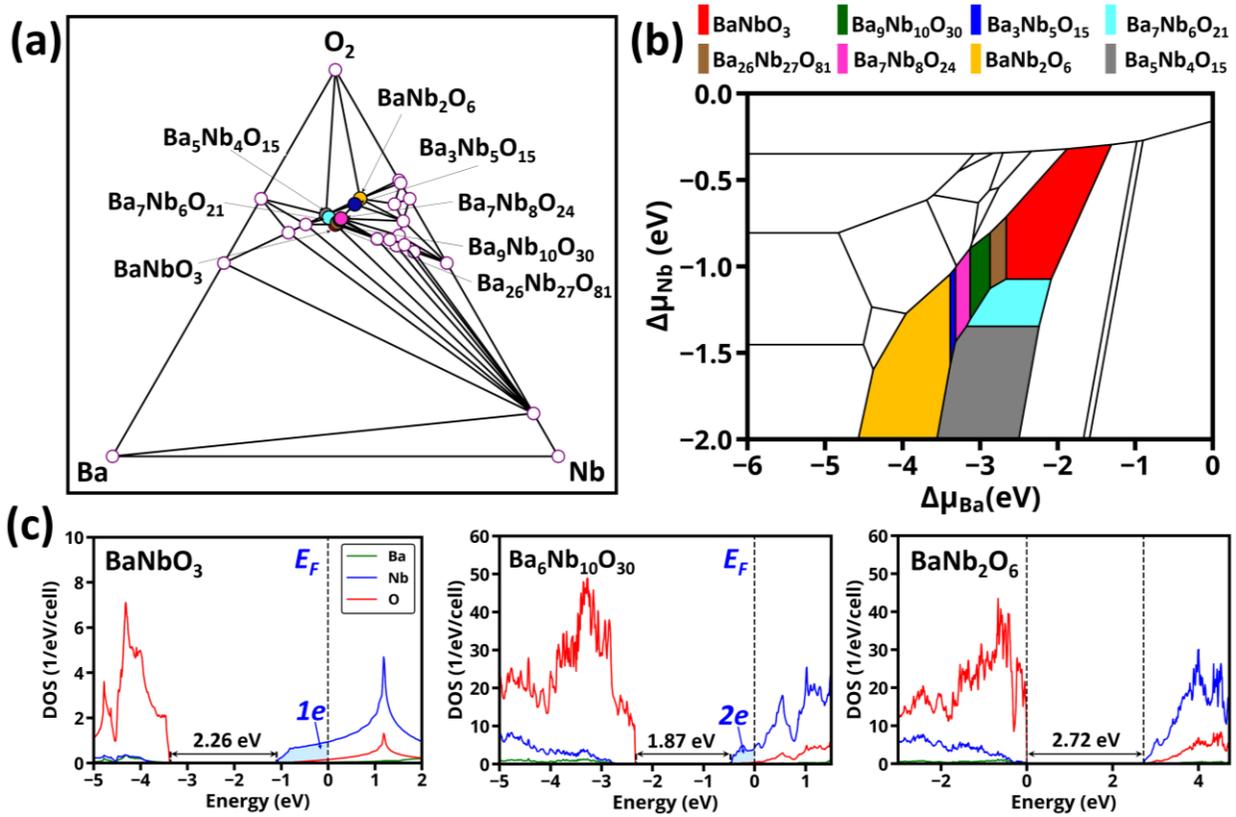

**Figure 4.** Stability and electronic properties of degenerate insulators and ordered vacancy compounds (OVCs) in the Ba-Nb-O system. (a) Convex hull for the Ba-Nb-O system indicating in color the OVC phases. (b) Chemical potential diagram for the Ba-Nb-O system, showing stability chemical potential zone for each stable OVC phase in colors corresponding to the ground states in (a). (c) Density of states (DOS) for $BaNbO_3$ and its OVCs, indicating the number of electrons in the conduction band per unit formula. In Fig.2b, the OVCs were not included in the calculations of stability green zone of $BaNbO_3$, but they are included in Figs. 4b. This redefines the chemical potential stability zone under which the base compound can exist and consequently the formation energy of Ba vacancies in the base compound is about 0 eV under the redefined cation-poor conditions.

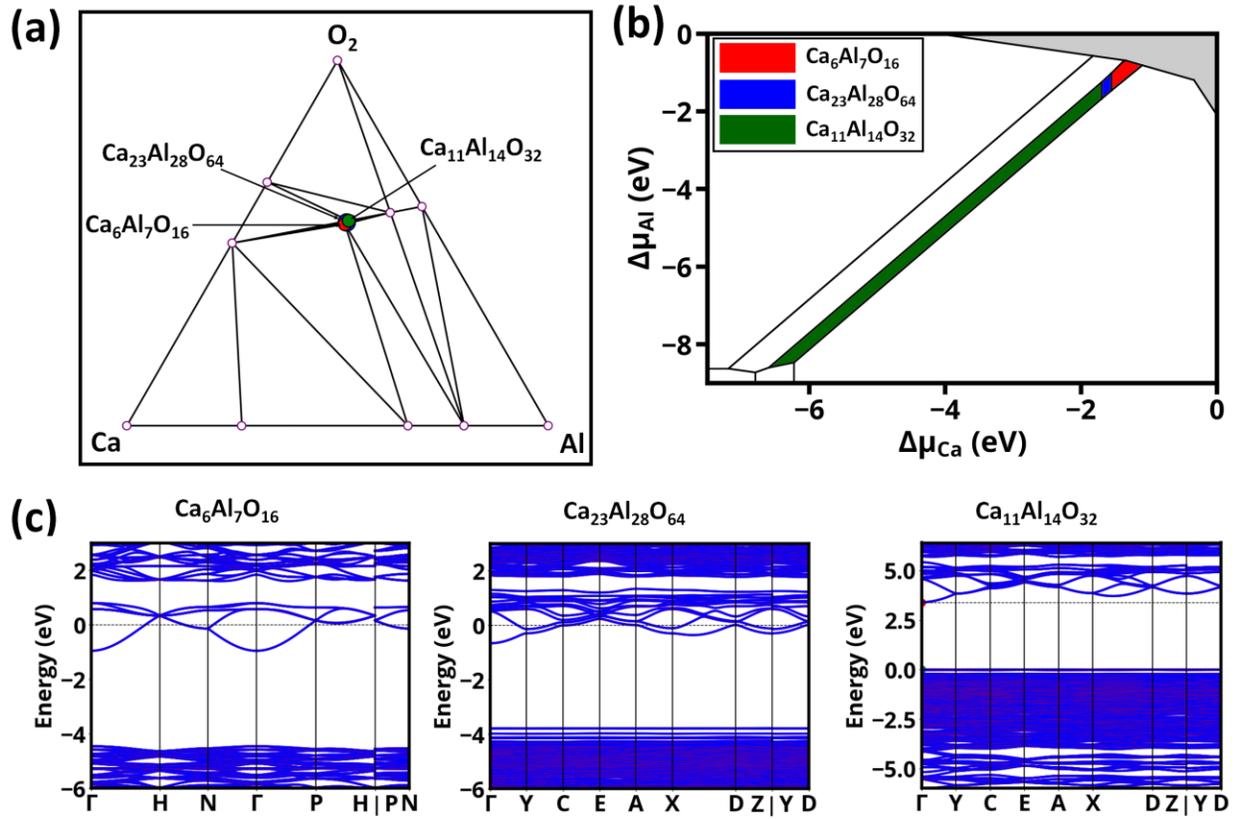

**Figure 5.** Stability and electronic properties of degenerate insulators and ordered vacancy compounds (OVCs) in the Ca-Al-O system. (a) Convex hull for the Ca-Al-O system indicating in color the OVC phases. (b) Chemical potential diagram for the Ca-Al-O system, showing stability chemical potential zone for each stable OVC phase in colors corresponding to the ground states in (a). Grey zone corresponds to prohibitive chemical potential-stability zone of binary Ca-Al systems. (c) Band structures for $Ca_6Al_7O_{16}$ and its ordered vacancy compounds. In Fig.2d, the OVCs were not included in the calculations of stability green zone of $Ca_6Al_7O_{16}$, but they are included in Figs. 5b. This redefines the chemical potential stability zone under which the base compound can exist and consequently the formation energy of Ca vacancies in the base compound is about 0 eV under the redefined cation-poor conditions.

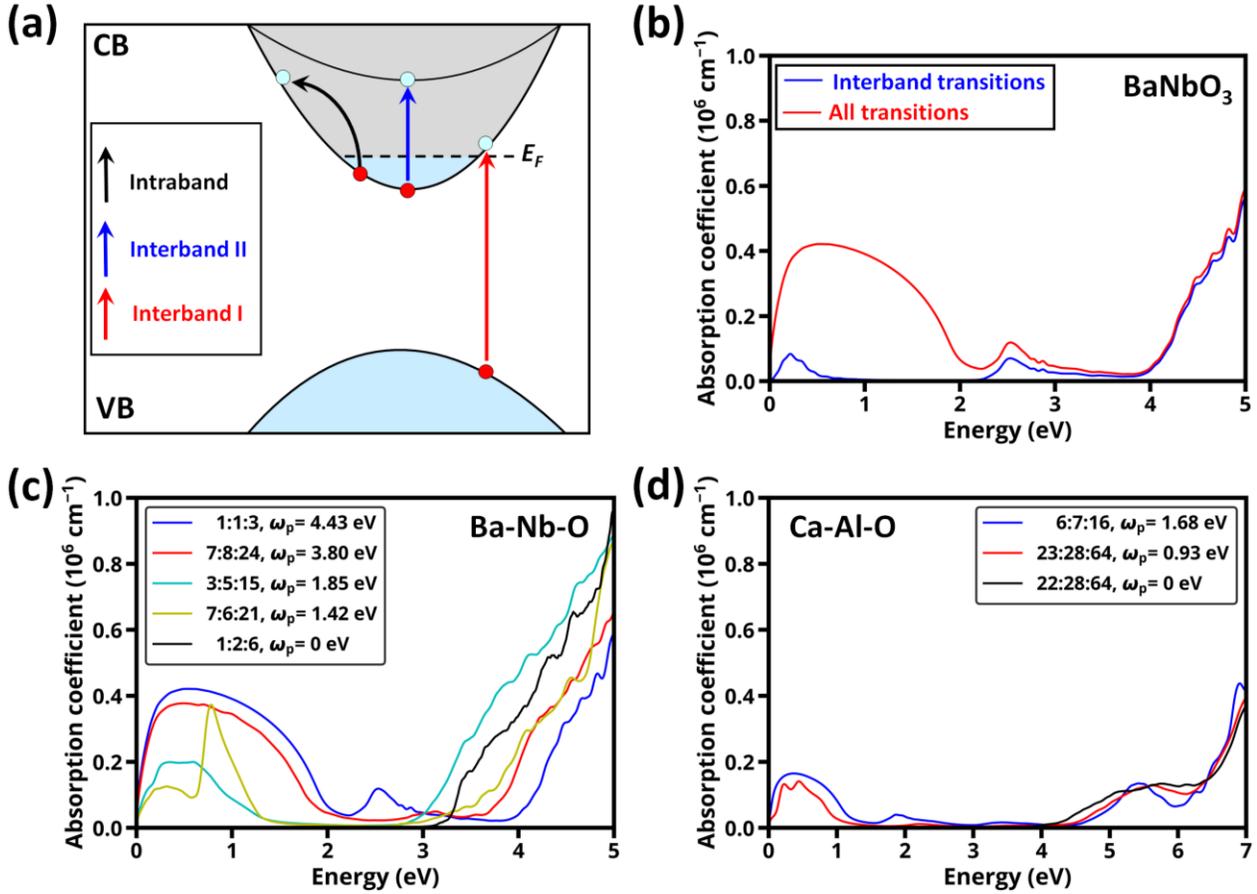

**Figure 6.** Effect of non-stoichiometry on optoelectronic properties. (a) Schematic illustration of different contributions to optical properties in degenerate insulators. (b) Absorption spectra for $BaNbO_3$ considering only interband transitions and superposition of interband and intraband transitions. (c) Effect of non-stoichiometry on average absorption spectra of degenerate insulators in the Ba-Nb-O system. (d) Effect of non-stoichiometry on average absorption spectra of degenerate insulators in the Ca-Al-O system. The averaged plasma frequency over the three Cartesian directions is given for each system.